\documentclass[twocolumn,english]{svjour3}
\usepackage[T1]{fontenc}
\usepackage[latin9]{inputenc}
\usepackage{color}
\usepackage{babel}
\usepackage{array}
\usepackage{url}
\usepackage{multirow}
\usepackage{amsmath}
\usepackage{graphicx}
\usepackage[numbers,sort&compress]{natbib}
\usepackage[unicode=true,pdfusetitle,
 bookmarks=true,bookmarksnumbered=false,bookmarksopen=false,
 breaklinks=false,pdfborder={0 0 0},backref=false,colorlinks=true]
 {hyperref}
\hypersetup{
 linkcolor=blue, urlcolor=blue, citecolor=blue, pdfstartview={FitH}, hyperfootnotes=false, unicode=true}

\makeatletter

\renewcommand{\fnum@figure}{\bf{Fig.}~\thefigure}
\usepackage{caption}
\usepackage{breqn}
\usepackage[misc,geometry]{ifsym}
\renewcommand\refname{References}
\usepackage{cite}

\makeatother

\begin{document}

% Journal Name %
\journalname{Applied Physics B}

\title{Direct observation of atomic diffusion in warm rubidium ensembles}

\author{Micha\l\ Parniak \and Wojciech Wasilewski}

\institute{Micha\l\ Parniak (\Letter) \and Wojciech Wasilewski (\Letter) \\Institute of Experimental Physics,
University of Warsaw, Ho\.{z}a 69, 00-681, Warsaw, Poland \\ e-mail: michal.parniak@student.uw.edu.pl \\ tel.: +48 697 555 351 \\ e-mail: wwasil@fuw.edu.pl \\ tel.: +48 22 55 32 120}
\maketitle
\begin{abstract}
We present a robust method for measuring diffusion coefficients of
warm atoms in buffer gases. Using optical pumping, we manipulate the
atomic spin in a thin cylinder inside the cell. Then we observe the
spatial spread of optically pumped atoms in time using a camera, which
allows us to determine the diffusion coefficient. As an example, we
demonstrate measurements of  diffusion coefficients of rubidium in
neon, krypton and xenon acting as buffer gases. We have determined
the normalized (273 K, 760 Torr) diffusion coefficients to be {0.18$\pm0.03$
cm$^{2}$/s for neon, 0.07$\pm0.01$ cm$^{2}$/s for krypton, and
0.052$\pm0.006$ cm$^{2}$/s for xenon. }
\end{abstract}

\section{Introduction}

Warm atomic ensembles have recently become a very useful tool in modern
quantum engineering. The most notable applications include quantum
memories \citep{VanderWal2003,Appel2008} and quantum repeaters \citep{Duan2001}
that can lead to development of quantum networks \citep{Kimble2008}.
Warm atoms have also been  used as a medium for four-wave mixing \citep{McCormick2008},
electromagnetically induced transparency (EIT) \citep{Fleischhauer2005}
and ultraprecise magnetometry \citep{Chalupczak2012}. 

While experiments with vapors contained in sealed cells are relatively
simple, they are  limited by inevitable atomic motion. Typically a
buffer gas is used to contain the atoms and make their motion diffusive.
In all of the above examples the diffusion rate was among primary
performance limiting factors. Its importance has been recognized and
its effect on EIT \citep{Shuker2008,Firstenberg2008,Firstenberg2010,Yankelev2013}
and on the Gradient Echo Memory \citep{Glorieux2012,Higginbottom2012,Luo2013}
have been studied both experimentally and theoretically. 

Prior knowledge of  diffusion coefficients enables designing optimal
experiments and greatly facilitates the interpretation of the results.
However, there is a striking lack of precise measurements of  diffusion
coefficients in various buffer gases. We believe that the reason for
this is unavailability of robust methods. In most cases  diffusion
coefficients were deduced using methods designed for studying spin-exchange
of optically aligned atoms \citep{Franzen1959}. 

The lack of both data and methods motivated us to develop a simple
and robust procedure designed specifically to measure the diffusion
coefficients. We demonstrate it on an example of  diffusion of rubidium
in neon, krypton and xenon. 

In our method, we  pump optically a thin pencil-shaped volume of atoms
inside a given cell using a short laser pulse. After the pump pulse
ends, we wait for a varying time and let rubidium atoms diffuse. Then
we send a pulse from a probe laser in a beam that covers nearly the
entire cell. The probe light is virtually unaffected by  pumped atoms
but absorbed by the unpumped ones. Therefore  spatial distribution
of the transmitted probe light reveals how far the atoms have travelled
between the pump and probe pules and thus provides the diffusion coefficient.

This paper is organized as follows. In section 2 we present a simple
model that describes our method. In section 3 we describe in detail
our experimental setup. In section 4 we present practical methods
for analyzing the data obtained. Section 5 gives the results of our
measurements and compares them with theoretical predictions as well
as with the results obtained previously. Section 6 concludes the paper.

\section{Method}

In out method, we register a decrease in optical depth of the atomic
sample due to optical pumping $\Delta OD(x,y,t)$ as a function of
spatial position $(x,y)$ and the delay $t$ between the pumping and
the actual observation. In practice, the difference $\Delta OD$ may
be computed by measuring the intensity of light that passes through
our cell with $I_{\mathrm{p}}(x,y,t)$ and without optical pumping
$I_{\mathrm{np}}(x,y)$. The formula reads:

\begin{equation}
\Delta OD(x,y,t)=\ln\left(\frac{{I_{\mathrm{p}}(x,y,t)}}{I_{\mathrm{np}}(x,y)}\right).
\end{equation}

The decrease in optical depth is also proportional to the decrease
in density of the atoms in the ground state of the atomic transition
to which the probe is coupled, that is $c_{\mathrm{np}}-c_{\mathrm{p}}(x,y,t)$,
where $c_{\mathrm{np}}$ stands for the equilibrium density observed
without pumping. Since the density $c_{\mathrm{p}}(x,y,t)$ at time
$t$ after pump pulse evolves according to the diffusion equation,
so does the decrease in the optical depth $\Delta OD(x,y,t)$. 

In addition to the diffusion, other mechanisms, such as spin-exchange
collisions, may urge the density $c_{\mathrm{p}}(x,y,t)$ towards
the steady state value $c_{\mathrm{np}}$. We call the rate of those
relaxation processes $\gamma_{0}$ and observe that it is position-independent.
Later we incorporate it into data analysis. Moreover, the atoms may
relax in collisions with cell walls. In our experiments the observation
times $t$ were too short for pumped atoms to reach the side walls.
Some portion of the initially pumped volume would reach cell windows,
however since the length of the cell is almost 100 times the typical
diffusion distance, the boundary effects are negligible.

It is convenient and intuitive to assume that the density $c_{\mathrm{p}}(x,y,t)$ after 
 pumping is $z$-independent, which requires that the pump beam should 
saturate the absorption in the ensemble. In fact, this assumption is not necessary, 
 as every solution to the 3D diffusion equation (in this case with additional relaxation) can be written as sum  
of separable solutions, that is $c_{\mathrm{np}}-c_{\mathrm{p}}(x,y,z,t)=\sum_j u_{j}(x,y,t)f_{j}(z,t) \exp(-\gamma_0 t)$, where both $u_{j}(x,y,t)$ and $f_{j}(z,t)$ satisfy the diffusion equation in respective coordinates. 
Now we integrate over $z$ to finally obtain the decrease in the optical depth $\Delta OD(x,y,t)$, but the integral $\int_0^L f_{j}(z,t) \mathrm{d}z$ is time-independent, as the relaxation at optical windows is negligible. The optical depth difference will satisfy the 2D diffusion equation with relaxation, as it is now a linear combination of $u_{j}(x,y,t)$ with time-independent coefficients. 

\begin{figure}
\centering{}\includegraphics[bb=10bp 0bp 363bp 220bp,clip,scale=0.7]{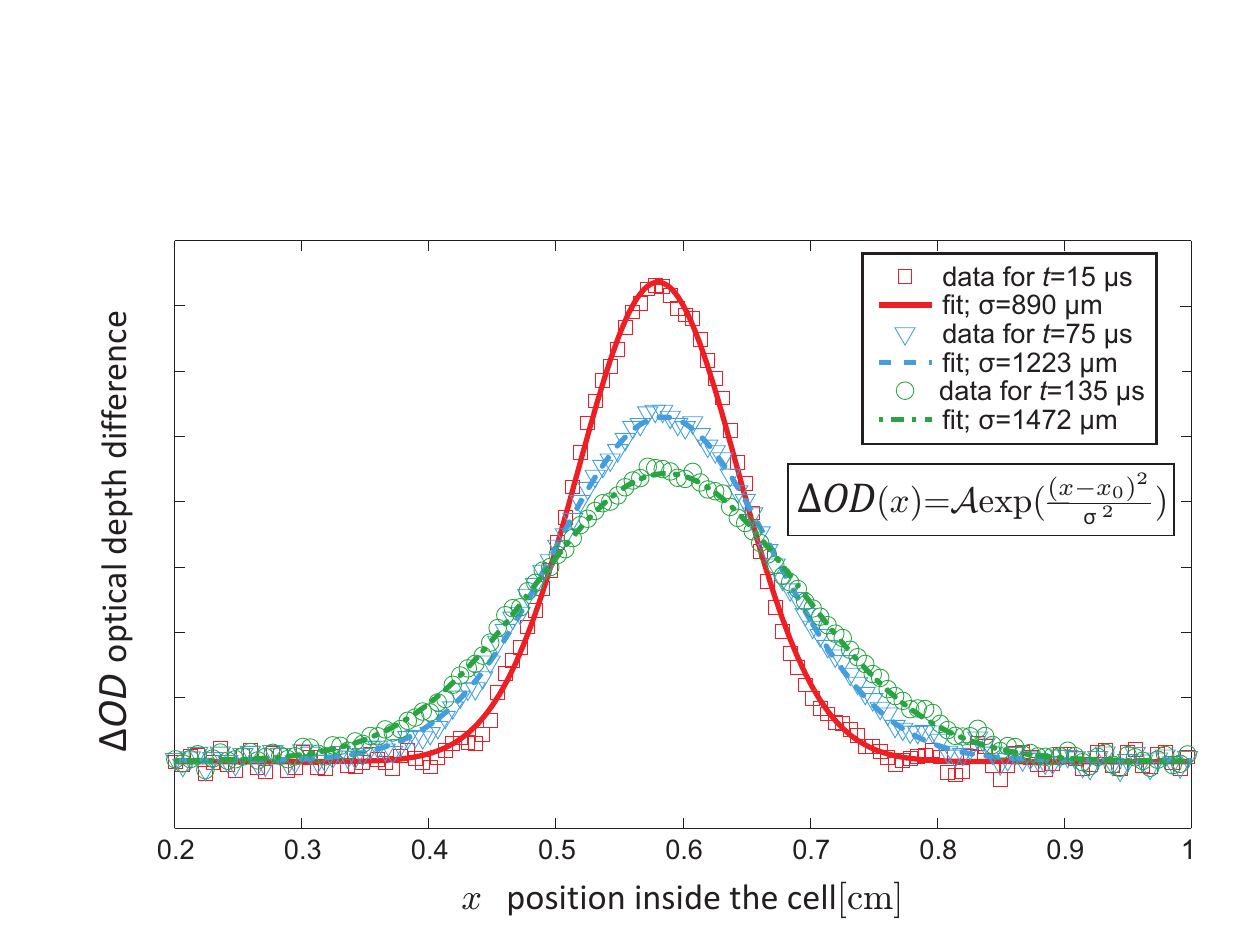}\caption[Obtained optical depth difference]{Typical examples of obtained optical depth difference $\Delta OD(x,y=0,t)$
maps at various times $t$ after pump pulse.  Gaussian fits give us
widths of these distributions $\sigma_{x}(t)$ as a function of  diffusion
time.\label{fig:gaussy}}
\end{figure}

By fitting Gaussians to obtained $\Delta OD(x,y,t)$ cross-sections
as presented in Fig. \ref{fig:gaussy}, we can estimate the diffusion
coefficient $D$ using the diffusion rule for the width of fitted
Gaussians $\sigma_{x}(t)=\sqrt{{\sigma_{x}(t=0)^{2}+4Dt}}$. A more
universal method is presented in Data Analysis section.

\section{Experimental Setup}

\begin{figure*}
\centering{}\includegraphics[bb=0bp 100bp 750bp 595bp,clip,scale=0.45]{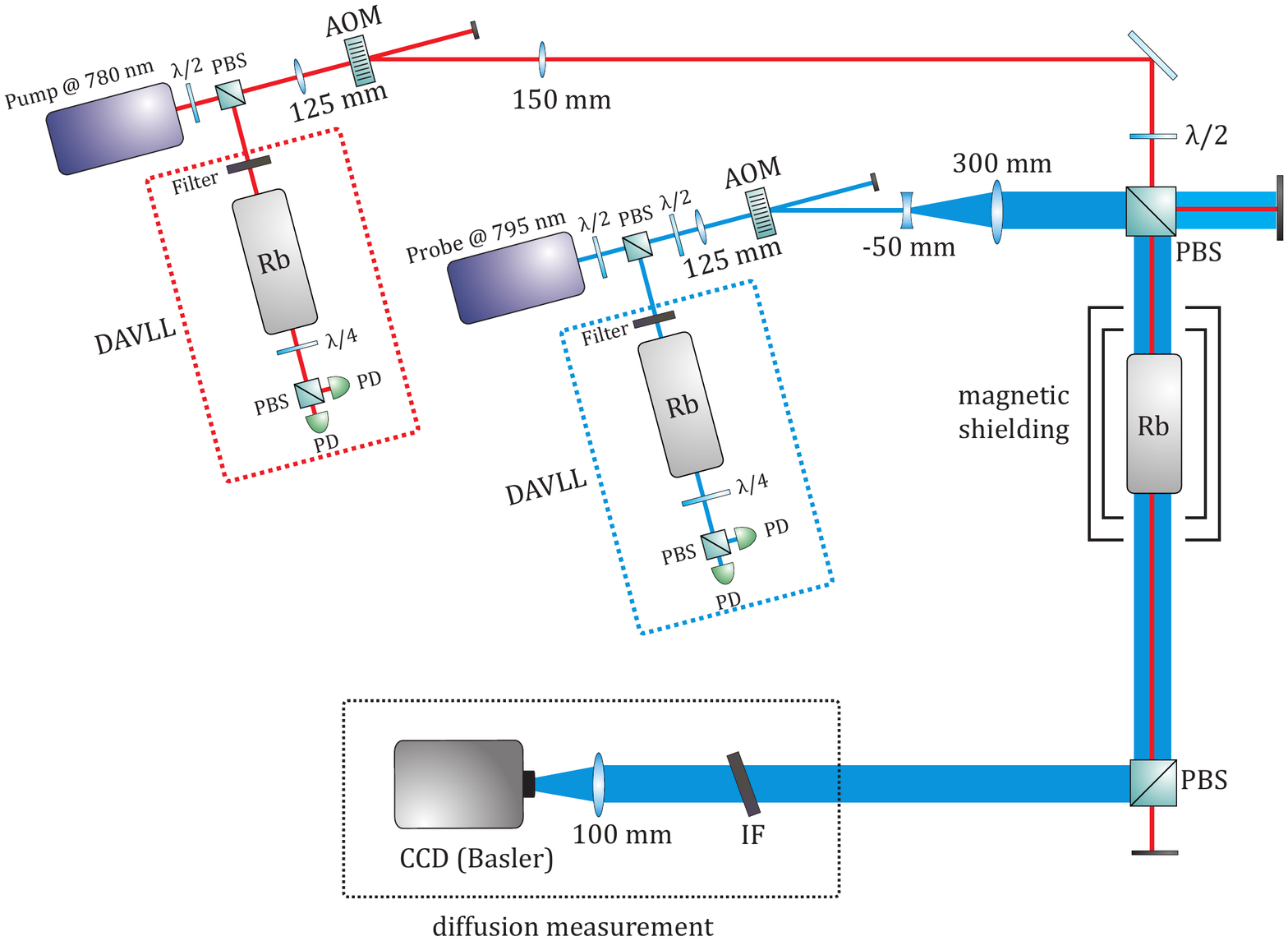}\caption[Schematic of our experimental setup]{Schematic of our experimental setup; \textit{AOM} acousto-optic modulator,
\textit{DAVLL} dichroic atomic vapor laser lock setup {\citep{KLCorwinZTLuCFHandRJEpsteinCEWieman1998}},
\textit{PD} photodiode, \textit{PBS} polarizing beam-splitter, \textit{Rb}
rubidium vapor cell, \textit{$\lambda/2$} half-wave plate, \textit{$\lambda/4$}
quarter-wave plate, \textit{IF} interference filter, \textit{CCD}
charge coupled device camera. Lenses are labeled with their focal
lengths.\label{fig:setup-1}}
\end{figure*}

Fig. \ref{fig:setup-1} presents a schematic of the experimental setup.
Both pump and probe lasers are Toptica distributed feedback laser
diodes, each frequency stabilized using dichroic atomic vapor spectroscopy
\citep{KLCorwinZTLuCFHandRJEpsteinCEWieman1998}.

In case of rubidium 87 the pump laser was tuned to {$F_{g}=1\rightarrow F_{e}=0,1,2$}
transitions on D2 line and the probe laser to {$F_{g}=1\rightarrow F_{e}=2$}
transition on D1 line. In case of rubidium 85 the pump laser was tuned
to {$F_{g}=3\rightarrow F_{e}=2,3,4$}
transitions on D2 line and the probe laser to {$F_{g}=3\rightarrow F_{e}=2,3$}
transitions on D1 line. We have found that the stability of the pump
laser is important, as the pumping rate needs to be constant, although
the laser does not need to be tuned precisely to the center of any
transition.

The laser beams are directed onto the acousto-optic modulators (AOM)
used to produce the pulse sequence. The beam shaping optics follows.
Both beams are collimated and the probe beam is expanded to a desired
$1/e^{2}$ waist diameter of about 1 cm, while the pump beam has a
waist diameter of 1 mm. 

The beams are joined on a polarizing beam splitter (PBS). Half-wave
plates before the joining point allow us to  control the power of
both beams precisely. After the PBS, the beams are parallel and overlap.

Both beams pass through a quartz cell (Precision Glassblowing, 25
mm in diameter, various lengths) filled with isotopically pure rubidium vapor and  buffer
gas. The vapor cell is placed inside a $\mu$-metal magnetic shielding
to avoid pumping alternations due to external magnetic filed. The
cell is mounted in flexible aluminum sleeves heated with a bifilar-wound
copper coil. Despite the bifilar winding, we have found that it is
better to stop the heating for the time of measurement to avoid disturbing
the pumping. The holder temperature stabilization is based on the
resistance of the coil's windings. 

The vapor temperature is determined by measuring  absorption spectrum
of the cell and fitting the result with a theoretical curve. We keep
the temperature within the $5^{\circ}$C range around $40^{\circ}$C.

After passing through the cell, the pump beam is filtered out by
a PBS and an interference filter. We image the inside of the cell
on a CCD camera (Basler{,
scA1400-17fm}) with a single lens. Magnification
of this optical system was both calculated and measured with a cell
replaced by a reference target. These two methods led to consistent
results, showing that camera's pixel pitch correspond to a $59\pm2$
$\mu$m distance inside the cell. 

The camera is triggered synchronously with laser pulses. We use minimal
shutter duration (40 $\mu$s) to minimize the background coming mainly
from scattering on the AOMs. In addition, by controlling the time
when shutter opens, we minimize the amount of pump light registered
by the camera. We achieved a relatively low background level, dominated
by the electronic offset. It was sufficient to subtract constant background
intensity from each image frame.

\begin{figure}
\begin{centering}
\includegraphics[scale=0.65]{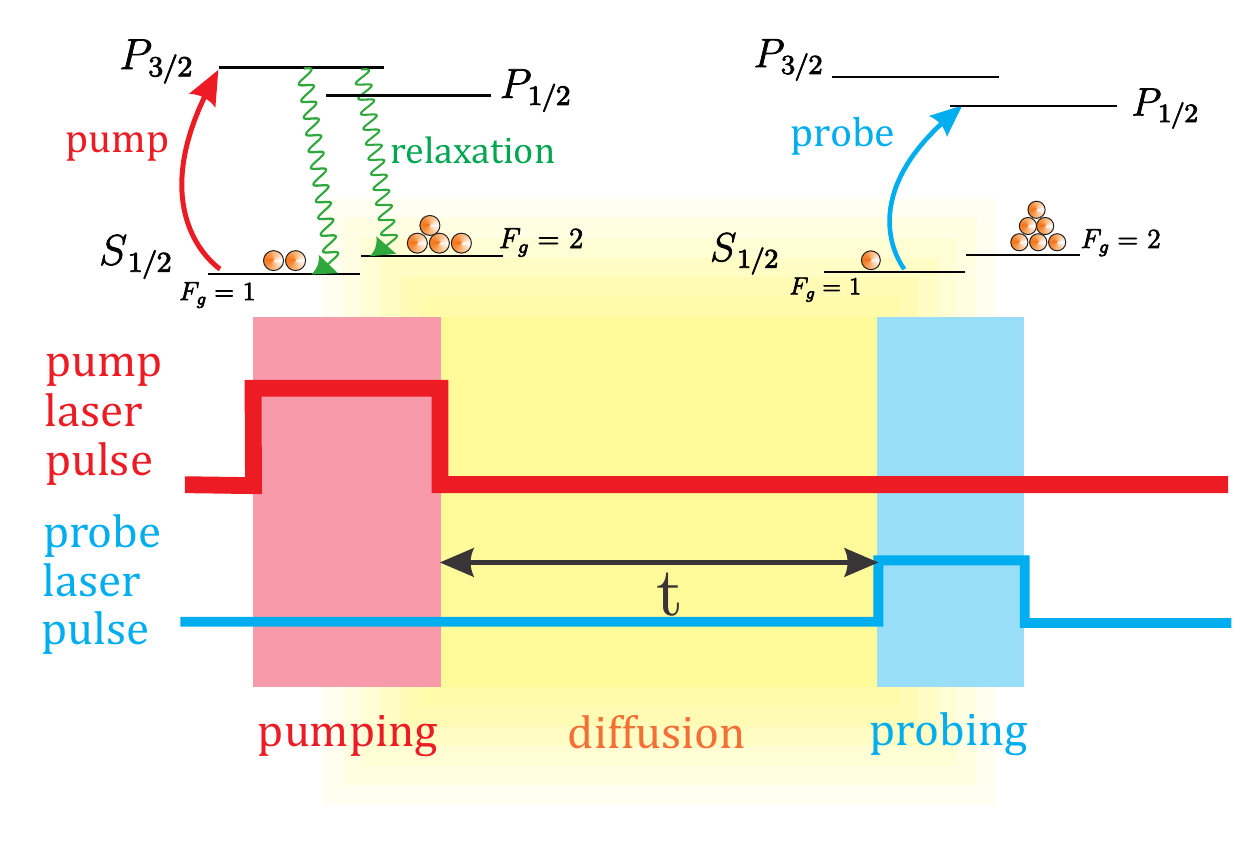}
\par\end{centering}

\caption[Light pulse sequence used in the experiment]{Light pulse sequence we use in our experiment. The top line represents
the pump laser tuned to rubidium D2 line and the bottom line represents
the probe laser tuned to D1 line. On the energy level diagrams (in
this case of rubidium 87 atom) we show how the optical pumping influences
populations: during pumping on the left and during probing on the
right. Wavy arrows represent relaxation mainly due to spontaneous
emission. }

\label{pulses}
\end{figure}

The pulse sequence is represented schematically in Fig. \ref{pulses}.
To measure the reference intensity of the probe light $I_{\mathrm{np}}(x,y)$,
we send a $0.4$ $\mu$s probe pulse alone and image the beam shape
after passing through the cell. To measure the probe light intensity
with the optical pumping present $I_{\mathrm{p}}(x,y,t)$, first we
apply a $0.5$ $\mu$s pump pulse and then wait for a varying time
$t$ and apply a $0.4$ $\mu$s probe pulse. The initial $\Delta OD(x,y,t=0)$ is diffused only slightly due to the short pump pulse. The unsaturated absorption of the pump light was about 70\%, but saturation effects caused half of the pump pulse to be transmitted through the cell. It resulted in optical pumping being weakly dependent on $z$.
The delay time $t$ varies from 1 to 150 $\mu$s. Each collected image of the probe beam is averaged
over 50 measurements, and the entire collection sequence for all delays
$t$ is repeated 10 times. This allows us to average over both long-term
(min) and short-term (ms) fluctuations.
The data collection rate is
limited by the 40 Hz frame rate of the camera.

\section{Data Analysis}

The initial shape of the pumped region is not easily described analytically,
therefore we have elected to use a data analysis method that can deal
with arbitrary functions. It is based on solving the diffusion equation
in the spatial-frequency Fourier domain, which is more general than
a Gaussian fit from Fig. \ref{fig:gaussy}. A similar method has previously
been  used in studies of fluorescence redistribution after photobleaching
\citep{Tsay1991}. Let us define the spatial Fourier transform of
the decrease of optical depths of the atomic sample due to optical
pumping $\Delta OD(x,y,t)$ as: 
\begin{multline}
\mathcal{{F}}\{\Delta OD\}(k_{x},k_{y},t)=\\
=\frac{{1}}{2\pi}\iint\mathrm{d}x\mathrm{d}y\Delta{OD}(x,y,t)\exp(-ik_{x}x-ik_{y}y).
\end{multline}

As we discussed in Method section the $z$-dependence of atomic density can be separated and $\Delta OD(x,y,t)$ becomes a linear combination of functions satisfying 2D diffusion equation with relaxation, so the time evolution of this Fourier transform due to diffusion takes
on a simple form

\begin{multline}
\mathcal{\mathcal{{F}}}\{\Delta OD\}(k_{x},k_{y},t)=\\
=\mathcal{{F}}\{\Delta OD\}(k_{x},k_{y},0)\exp(-(\underbrace{{\gamma_{0}+D(k_{x}^{2}+k_{y}^{2})}}_{\gamma({\bf k})})t).\label{eq:fourier}
\end{multline}

This solution tells us that a component of optical depth difference
having a certain spatial periodicity given by $k_{x}$ and $k_{y}$
(that may be also called spatial frequencies) decays exponentially
at a rate 
\begin{equation}
\gamma(k_{x},k_{y})=\gamma_{0}+D(k_{x}^{2}+k_{y}^{2}).\label{eq:gamma_k}
\end{equation}

The procedure of data analysis according to the above equations is
as follows. Having measured optical depths differences $\Delta OD(x,y,t)$
for various delay times $t$, we perform a numerical Fourier transform
of each map and get $\mathcal{\mathcal{{F}}}\{\Delta OD\}(k_{x},k_{y},t)$
for different delay times $t$.

\begin{figure}
\begin{centering}
\includegraphics[scale=0.65]{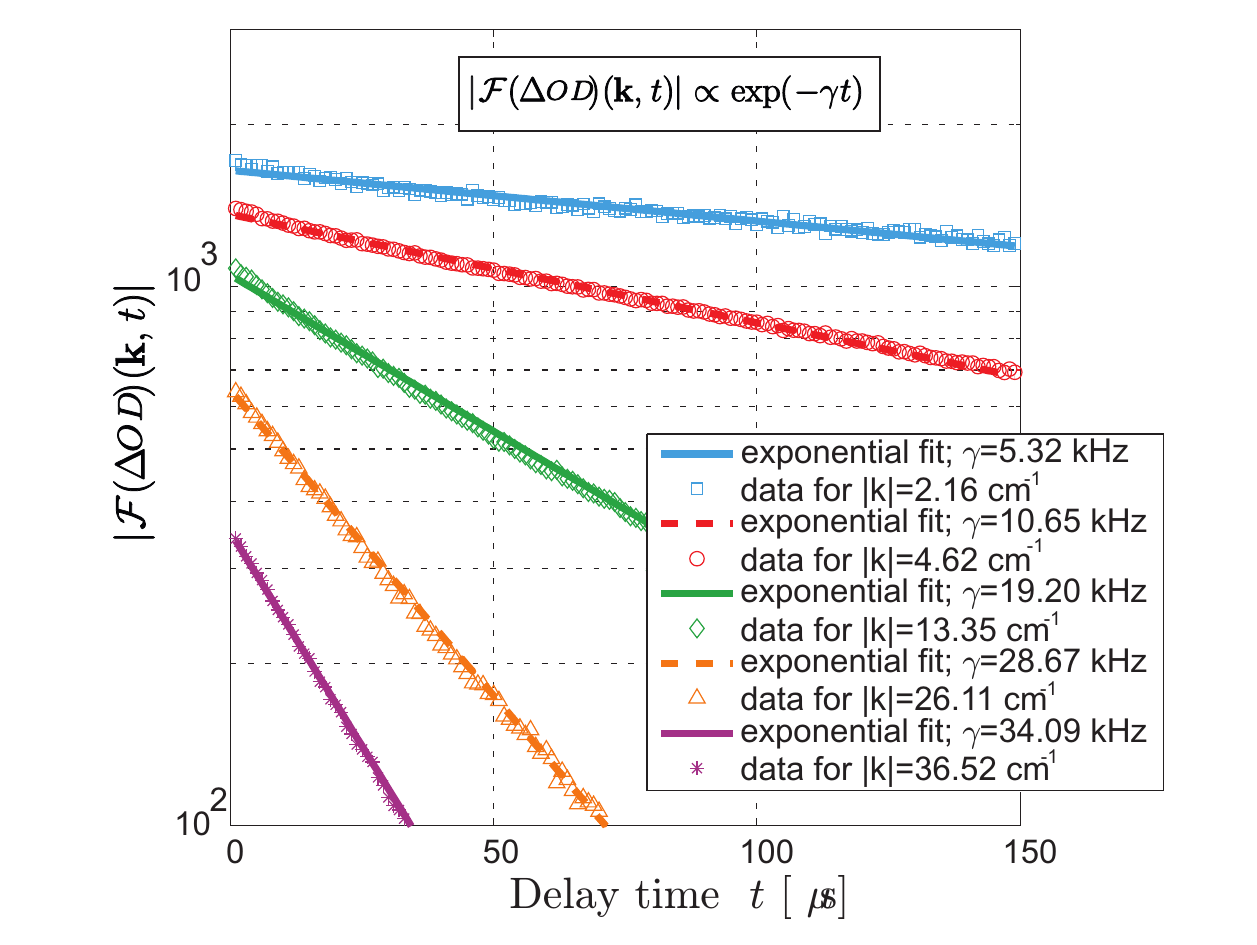}
\end{centering}

\centering{}\caption[Examplary exponential decays]{Examplary exponential decays of $\mathcal{{F}}\{\Delta OD\}(|\mathbf{k}|,t)$
with time between pump and probe, for various components with spatial
wavevectors$\mathbf{k}$ in the image of optical depth decrease $\Delta{OD}(x,y,t)$.
Data for $^{85}$Rb cell with 5 Torr of neon.\label{fig:decays} }
\end{figure}

\begin{figure}
\centering{}\includegraphics[bb=115bp 264bp 750bp 555bp,clip,scale=0.65]{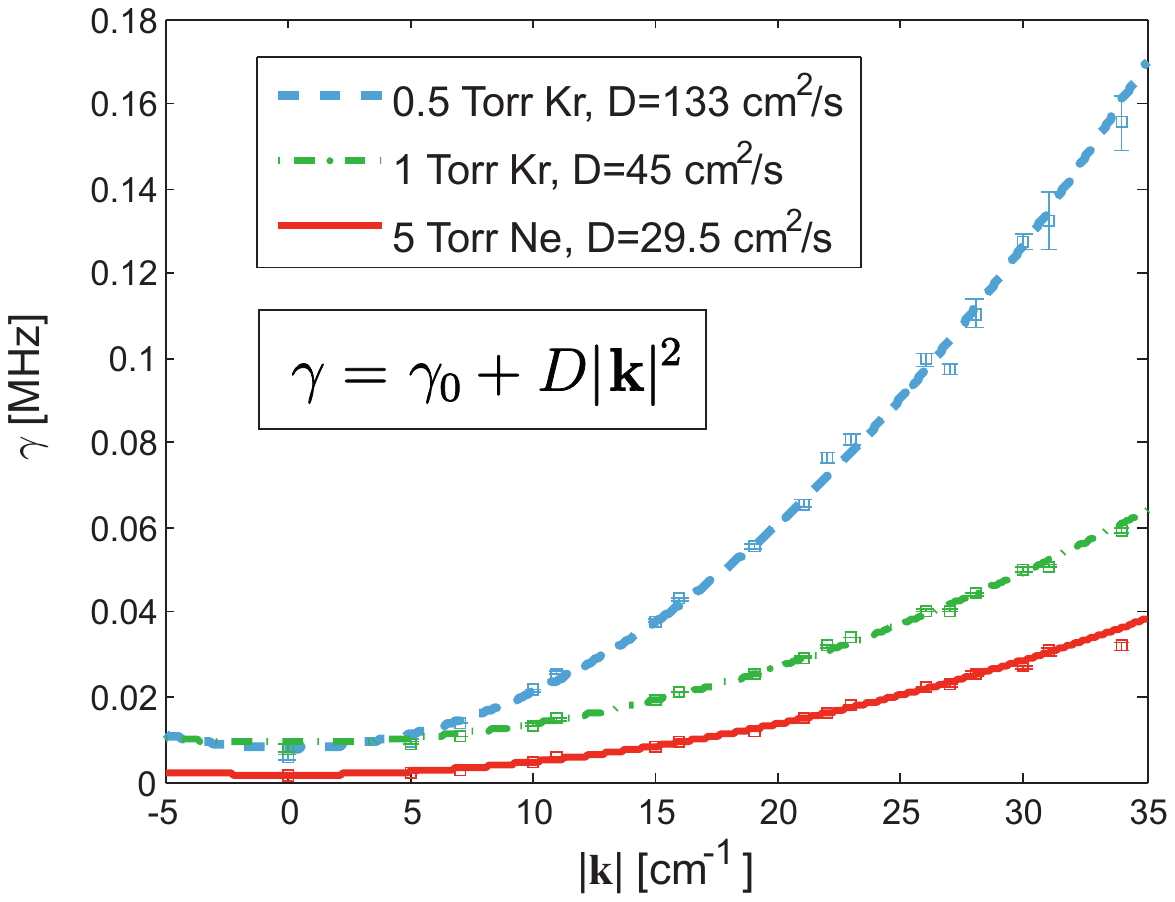}\caption[Quadratic fits of the decay rate $\gamma(|\mathbf{k}|)$]{Quadratic fits of the decay rate $\gamma(|\mathbf{k}|)$ as a function
of spatial wavevector $\mathbf{k}$ to the data obtained in the Fourier
procedure of data analysis, as in Fig.\ref{fig:decays}.\label{fig:quadratic} }
\end{figure}

Next, for each $\mathbf{k}=(k_{x,}k_{y})$ we fit a simple exponential
decay model to the absolute amplitude of a component of the optical
depth decrease with a given spatial periodicity. In practice this
is possible only for components of high enough amplitudes, that is
for $|\mathbf{k}|$ smaller than the inverse pump beam width. The
fit result is the decay rate as a function of spatial periodicity
$\gamma(\mathbf{k})$. Examplary decay fits for various spatial wavevectors\textbf{
$\mathbf{k}$} are presented in Fig. \ref{fig:decays}. As this decay
rate should only depend on the length of \textbf{$\mathbf{k}$} vector,
we perform an angular averaging procedure to obtain a one-variable
function $\gamma(|\mathbf{k}|)$. Finally, we use the relation for
the decay rate $\gamma(|\mathbf{k}|)=\gamma_{0}+D|\mathbf{k}|^{2}$
that comes from the diffusion equation. We fit a parabola to the previously
computed $\gamma(|\mathbf{k}|)$ dependence and retrieve the diffusion
coefficient $D$. Examplary fits for various Rb cells we used are presented
in Fig. \ref{fig:quadratic}.

\section{Results}

\begin{table*}
\caption[Diffusion coefficients]{Measured diffusion coefficients, normalized diffusion coefficients
obtained from our measurements, theoretical predictions based on the
Chapman-Enskog formula, and some previous results of the diffusion
coefficients measurements.\label{table}}

\centering{}%
\begin{tabular}{ccc>{\centering}p{2cm}>{\raggedright}p{2cm}>{\raggedright}p{3.2cm}}
\hline 
Cell (buffer gas, Rb isotope)%
 & $D$ {[}cm$^{2}$/s{]} (40$^{\circ}$C) & \multicolumn{2}{c}{$D_{\circ}$ (this paper) {[}cm$^{2}$/s{]}} & $D_{\circ}$ (theory) {[}cm$^{2}$/s{]} & $D_{\circ}$ (previous results) {[}cm$^{2}$/s{]}\tabularnewline
\hline 
Ne 5 Torr, $^{85}$Rb & $29.5\pm1.0$ & 0.18$\pm$0.02 &\multirow{4}{2cm}{0.18$\pm0.03$} & \multirow{4}{2cm}{0.145$\pm0.01$} & \multirow{4}{3.2cm}{0.2 \citep{Chrapkiewicz2013}, 0.11 \citep{Shuker2008}, 0.18 \citep{Vanier1974},
0.31 \citep{Franzen1959,Gases1961}, 0.48 \citep{Franz1965}}\tabularnewline
Ne 2 Torr, $^{87}$Rb, paraf. & $62.0\pm0.8$ &0.15$\pm$0.02 & &  & \tabularnewline
Ne 100 Torr, $^{85}$Rb & $1.69\pm0.04$ & 0.21$\pm$0.03& &  & \tabularnewline
Ne 50 Torr, $^{87}$Rb & $2.70\pm0.04$ & 0.17$\pm$0.02& &  & \tabularnewline
\hline 
Kr 1 Torr, $^{87}$Rb & $71\pm2$ & 0.08$\pm$0.01 &\multirow{3}{2cm}{0.07$\pm0.01$} & \multirow{3}{2cm}{0.065$\pm0.005$} & \multirow{3}{3.2cm}{0.068 \citep{Chrapkiewicz2013}, 0.1 \citep{Bouchiat1972}, 0.04 \citep{Higginbottom2012}}\tabularnewline
Kr 1 Torr, $^{87}$Rb, paraf. & $45\pm2$ & 0.06$\pm$0.01& &  & \tabularnewline
Kr 0.5 Torr, $^{87}$Rb & $133\pm4$ &  0.08$\pm$0.01&  & & \tabularnewline
\hline 
Xe 1 Torr, $^{87}$Rb, paraf. & $43\pm1$ & \multicolumn{2}{c}{0.052$\pm0.006$} & 0.055$\pm0.005$ & 0.057 \citep{Chrapkiewicz2013}\tabularnewline
\hline 
\end{tabular}%
\end{table*}

The presented results were obtained using the Fourier method described
above. Special care was taken to ensure validity of each step, including
verification of rotational symmetry of $\gamma(\mathbf{k})$, which
allowed us to proceed with angular averaging to obtain $\gamma(|\mathbf{k}|)$. We noticed that good alignement was critical when ensuring that the spread of values of $\gamma(\mathbf{k})$ for a given $|\mathbf{k}|$ was small. Optimizing the setup resulted in this spread being much smaller than the uncertainty coming from each exponential fit, but both error sources were taken into account when calculating uncertainties.

\begin{figure}
\centering{}\includegraphics[bb=100bp 244bp 750bp 575bp,clip,scale=0.65]{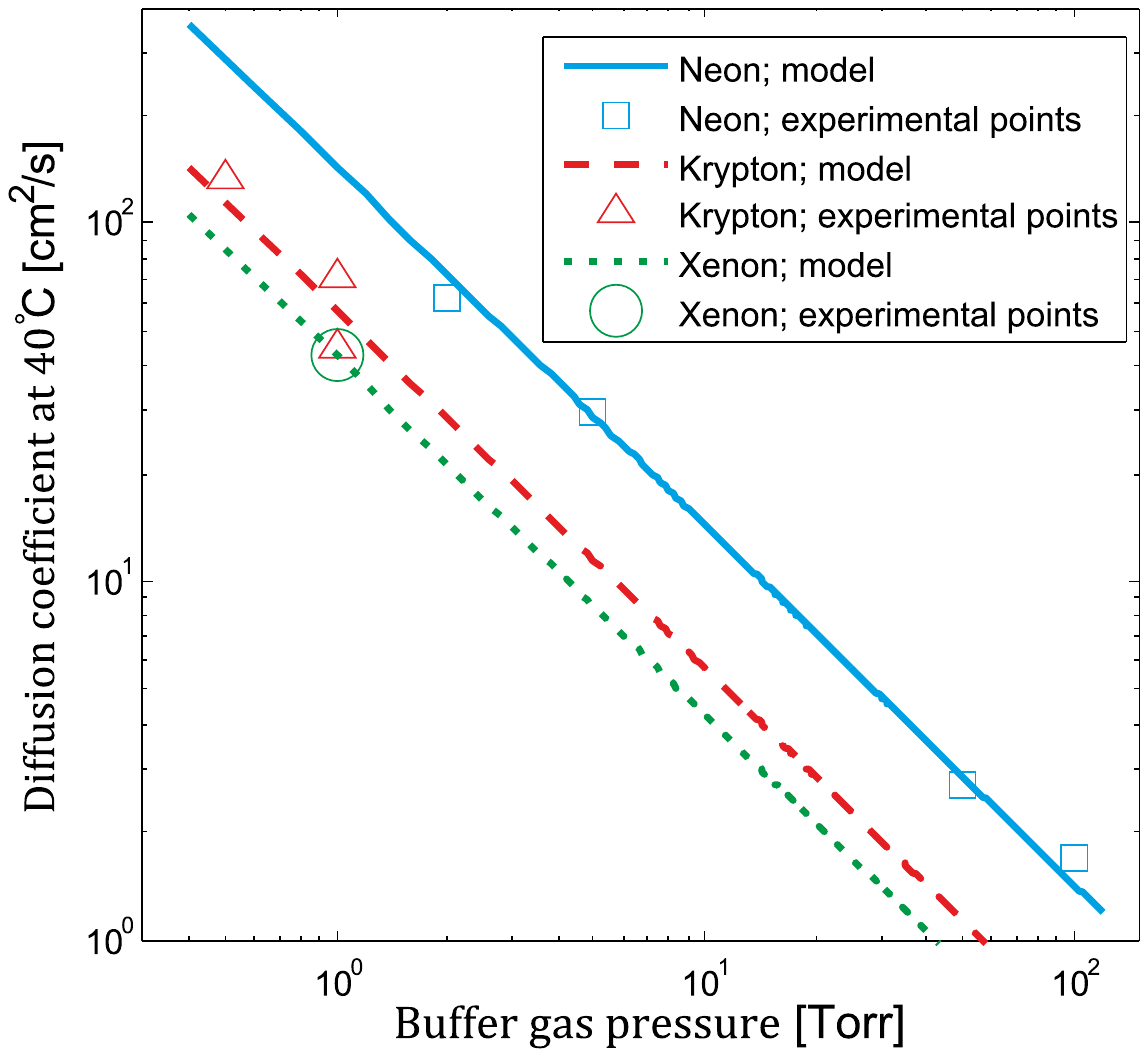}\caption[Summary of obtained diffusion coefficients]{Summary of obtained diffusion coefficients of rubidium in various buffer gases compared with inverse relation to buffer gas pressure predicted by the Chapman-Enskog model.\label{fig:summary}}
\end{figure}

Table \ref{table} ($D$, second column) contains diffusion coefficients
at 40$^{\circ}$C of rubidium in various cells, containing neon, krypton
or xenon. In the Fig. \ref{fig:summary} we see that the diffusion coefficient scales with inverse
of vapor gas pressure as expected \citep{Cussler1997} over a broad
range of buffer gas pressures, especially for neon. In this figure we also include curves representing model behaviour. The diffusion coefficients we measured in
two cells with 1 Torr krypton differ much more than expected. We believe
that this could be due to different true quantity of gas contained
in those cells due to variations  in the seal-off process. Most likely
the pressure was measured at significantly different temperatures
of the cell bodies. Another reason could be significant evaporation
of the paraffin coating which was present only in one cell as chemicals such as volatile hydrocarbons present in the coating could slow down the diffusion when in gaseous state.

To our knowledge, our group has been the first one to  measure directly
the diffusion coefficient of rubidium in xenon. The result we present
here confirms the previously measured value {\citep{Chrapkiewicz2013}}.

We scaled the results to normal conditions (760 Torr pressure and
$0^{\circ}$C temperature) for Ne, Kr and Xe using Chapman-Enskog
formula {\citep{Cussler1997}}.
We inferred uncertainties of normalized diffusion coefficient for neon and krypton from the spread of experimental results. In case of xenon we attribute the uncertainty of the result to the uncertainty of buffer gas pressure. Note that unnormalized diffusion coefficient for each cell are measured much more accurately.
Values of collision integrals at different temperatures were obtained
according to the procedure described in Ref. {\citep{Cussler1997}}.
Required collisional parameters were taken from Refs.{
\citep{Ghatee2007}} and {\citep{Gibble1991}}.
The ideal gas model was used for the relation between pressure of
the buffer gas and temperature. We believe that using the Chapman-Enskog
formula and normalized results we present here, it is possible to
calculate the diffusion coefficient of rubidium in Ne, Kr or Xe at
an arbitrary pressure or temperature. 

In the Table \ref{table} we compare our normalized experimental results
($D_{\circ}$, third column) with theoretical predictions based on
Chapman-Enskog formula and studies of atomic collisions ($D_{\circ}$,
fourth column). Our results display satisfactory conformity with theoretical
predictions. Note that both the normalization we perform and the theoretical
predictions are based on a fundamentally approximate theory, as the
Chapman-Enskog formula comes from an approximate solution of the Boltzmann
equation and requires various parameters that were calculated indirectly.

The results we present here differ significantly from some of the
previous results (last column in table \ref{table}), however we believe
ours are more reliable for several reasons. Firstly, the results of
this paper agree with the results we obtained using the same cells
{\citep{Chrapkiewicz2013}}
but a completely different method. Secondly, the former methods relied
on measuring the relaxation of the atomic spin alignments a function
of gas pressure to retrieve both the diffusion coefficient and the
spin-exchange rates \citep{Franzen1959}.

\section{Conclusions}

We have presented a robust and simple method for measuring the diffusion
coefficients of warm atoms in buffer gases. Our method might be used
in all systems where the phenomenon of optical pumping occurs. We
have shown that observing the spread of a optically pumped region
by applying a pulsed probe beam and imaging it on a camera is sufficient
to determine the diffusion coefficient of the system. The data analysis
involved Fourier transforming of measured optical depth differences
and finding  decay rates for components of different spatial periodicity.
This approach is robust and provides opportunities for data consistency
checks at various calculation stages.

Our method can be easily used to measure diffusion in countless physical
systems and is easy to implement. In principle one laser could be
sufficient for both pump and probe pulses.

As a demonstration we have measured the diffusion coefficients of
rubidium in neon, krypton and xenon. Our results are consistent with
both  theoretical predictions and  previous results. We have also
found the presented method very useful when it comes to characterizing
various sealed cells with rubidium and a buffer gas. Notably, it is
capable of capturing wide range of diffusion coefficients, of at least
two orders of magnitude.

We believe that our measurements will enable greater control and better
design of experiments with warm rubidium ensembles. In particular
we note that krypton and xenon appear to be very good buffer gases
for modern quantum applications; yet,  they have scarcely been  used
so far. Apart from us only one group has utilized krypton \citep{Hosseini2009},
while xenon has been used only by us \citep{Chrapkiewicz2013}. 

To our knowledge our group has been the first one to  measure actually
the diffusion coefficient of rubidium in xenon, with this paper confirming
Ref. \citep{Chrapkiewicz2013}. Given the recent applications of hyperpolarized
$^{129}$Xe in medical imaging, we hope that the precise value of
this diffusion coefficient will help optimize setups like the one
presented in Ref. \citep{Fink2005}. 
\begin{acknowledgements}
We acknowledge generous support from Konrad Banaszek and Rafa\l{}
Demkowicz-Dobrza\'{n}ski. This work was supported by the Foundation
for Polish Science TEAM project, EU European Regional Development
Fund and FP7 FET project Q-ESSENCE (Contract No. 248095), National
Science Center grant no. DEC-2011/03/D/ST2/01941 and by Polish NCBiR
under the ERA-NET CHIST-ERA project QUASAR.
\end{acknowledgements}

\bibliographystyle{bibstyle}
\bibliography{library}

\end{document}